\documentclass[aps,showpacs,preprintnumbers,amsmath,amssymb,nofootinbibt,referee]{revtex4}
\usepackage{graphicx}

\def\be{\begin{equation}}
\def\ee{\end{equation}}
\def\bea{\begin{eqnarray}}
\def\eea{\end{eqnarray}}

\begin{document}


\title{Cosmological anisotropy from non-comoving dark matter and dark energy}

\author{Tiberiu Harko$^1$}
\email{t.harko@ucl.ac.uk}
\author{Francisco S. N. Lobo$^2$}
\email{flobo@cii.fc.ul.pt}

\affiliation{$^1$Department of Mathematics, University College London, Gower Street,
London, WC1E 6BT, United Kingdom}
\affiliation{$^2$Centro de Astronomia e Astrof\'{\i}sica da
Universidade de Lisboa, Campo Grande, Edificio C8 1749-016 Lisboa,
Portugal}

\date{\today}

\begin{abstract}

We consider a cosmological model in which the two major fluid components of the Universe, dark energy and dark matter,   flow with distinct four-velocities. This cosmological configuration is equivalent to a single anisotropic fluid, expanding with a four-velocity that is an appropriate combination of the two fluid four-velocities. The energy density of the single cosmological fluid is larger than the sum of the energy densities of the two perfect fluids, i.e., dark energy and dark matter, respectively, and contains a correction term due to the anisotropy generated by the differences in the four-velocities. Furthermore, the gravitational field equations of the two-fluid anisotropic cosmological model are obtained for a Bianchi type I geometry. By assuming that the non-comoving motion of the dark energy and dark matter induces small perturbations in the homogeneous and isotropic Friedmann-Lemaitre-Robertson-Walker type cosmological background, and that the anisotropy parameter is small, the equations of the cosmological perturbations due to the non-comoving nature of the two major components  are obtained. The time evolution of the metric perturbations  is explicitly obtained for the cases of the exponential and power law background cosmological expansion. The imprints of a non-comoving dark energy - dark matter on the  Cosmic Microwave Background and on the luminosity distance are briefly discussed, and the temperature anisotropies and the quadrupole are explicitly obtained in terms of the metric perturbations of the flat background metric.
Therefore, if there is a slight difference between the four-velocities of the dark energy and dark matter, the Universe would acquire some anisotropic characteristics, and its geometry will deviate from the standard FLRW one. In fact, the recent Planck results show that the presence of an intrinsic large scale anisotropy in the Universe cannot be excluded {\it a priori}, so that the model presented in this work can be considered as a plausible and viable working hypothesis.

\end{abstract}

\maketitle

\section{Introduction}

The recently released Planck data of the 2.7 full sky survey \cite{1,2,3,4,5} have shown a number of intriguing features, whose explanation will certainly require a deep change in our standard view of the Universe. These recent observations have
 measured the cosmic microwave background (CMB) to an unprecedented precision. One of
the most striking results of the nominal mission is that the best-fit Hubble constant has the value $H_0 = 67.4\pm 1.2$ km s$^{-1}$Mpc$^{-1}$, with a dark energy density parameter
$\Omega _{\Lambda } = 0.686 \pm 0.020$, and a matter density parameter $
\Omega _M = 0.307 \pm 0.019$.  Even though generally the Planck data confirm the foundations of the $\Lambda $CDM ($\Lambda $Cold Dark Matter) model, the observational data show some tension between the fundamental principle of the model and observations. For example, the Planck data
combined with the WMAP polarization data show that the index of the power spectrum is given by $n_s=0.9603\pm0.0073 $ \cite{2,3}, at the pivot scale $k_0 = 0.05$ Mpc$^{-1}$, which rules out the exact
scale-invariance ($n_s = 1$) at more than $5\sigma $. In addition to this, the joint constraints on $r$ (tensor-to-scalar ratio) and $n_s$ can constrain inflation models. For example, inflationary models with a
power-law potential $\phi ^4$  cannot provide a reasonable number of
$e$-folds (between 50-60) in the restricted space of $r-n_s$ at
around a $2\sigma $ level.

More interestingly, the search for the background geometry and topology of the Universe reveals that a Bianchi pattern, corresponding to an isotropic geometry of the Universe may account, in a quite efficient way,  for some large-scale anomalies seen in the Planck data \cite{5}. A Bayesian search for an anisotropic Bianchi ${\rm VII_h}$ geometry was performed, by using the Planck data in \cite{5}.  In a non-physical setting, where the Bianchi parameters are decoupled from the standard cosmology, the observational data favour a Bianchi component with a Bayes factor of at least 1.5 units of log-evidence.  However, in the physically motivated setting where the Bianchi parameters are fitted simultaneously with the standard cosmological parameters, no evidence for a Bianchi ${\rm VII_h}$ cosmology was found \cite{5}.

According to the $\Lambda $CDM model, the material content of the Universe consists of two components: dark energy, and dark matter, respectively \cite{PeRa03}. The observed late-time cosmic acceleration of the Universe \cite{acc} can be successfully explained by introducing
either a fundamental cosmological constant $\Lambda $, which would represent an intrinsic curvature of space-time, or a dark energy, a hypothetical fluid component in the form of a zero-point-energy that pervades the whole Universe, which would mimic a cosmological constant (at least during the late stage of the cosmological evolution). Currently one of the main dark energy scenarios is based on the so-called {\it quintessence} \cite{quint}, where dark energy corresponds to a  scalar particle $\phi $. There are also other dark energy candidates, such as the Chaplygin gas model \cite{Chap}, in which dark energy is a hypothetical substance that satisfies an exotic equation of state in the form $p=A/\rho ^{-\alpha}$, where $p$ is the pressure, $\rho $  is the energy density, and $A$  and $\alpha $ are positive constants. Modifying gravity at large scales, or considering extra-dimensional cosmological models  can also provide an explanation of the late acceleration of the Universe \cite{Od, Mar1}.

Dark matter is thought to be composed of cold neutral weakly interacting massive particles, beyond those existing in the Standard Model of Particle Physics, and not yet detected in accelerators or in direct and indirect searches. There are many possible candidates for dark matter, the most popular ones being the axions and the weakly interacting massive particles (WIMP) (for a review of the particle physics aspects of dark matter see \cite{OvWe04}). Their interaction cross section with normal baryonic matter, while extremely small, are expected to be non-zero and we may expect to detect them directly.  Scalar fields or other long range coherent fields coupled to gravity, have been considered as potential dark matter candidates \cite{OvWe04}. Recently, the possibility that the galactic dynamics of massive test particles may be understood in the context of modified theories of gravity, without the need for dark matter, was also explored~\cite{dmmodgrav}.

Now, precise observations of the cosmic microwave background radiation  enable us to test the fundamental predictions of inflation on primordial fluctuations such as scale independence and Gaussianity \cite{1,2,3,4,5}. The statistical isotropy has been a robust prediction protected by the cosmic no-hair conjecture, which claims that inflation washes out classical anisotropy. However, several recent  observations of the large scale structure of the Universe have questioned the principles of homogeneity and isotropy \cite{an}. As combined with the recent Planck results, these robust observational results show that the presence of an intrinsic large scale anisotropy in the Universe cannot be excluded {\it a priori}. Recently, the possibility that the geometry of the Universe may not be of the Friedmann-Lemaitre-Robertson-Walker (FLRW) type was investigated from different points of view \cite{TK,an1}.  In particular, in \cite{TK} the homogeneous locally rotationally symmetric (LRS) class of metrics were considered as describing the global geometry of the Universe. Such spacetimes possess a preferred direction in the sky and at the same time a CMB which is isotropic at the background level. A distortion of the luminosity distances has been obtained, and used to test the model against the CMB and supernovae data. It has also been shown that the latter exhibit a higher-than-expected dependence on angular position. Local constraints from
limits from simulated distance measurements for various distributions of survey fields in a Bianchi I anisotropic universe were obtained \cite{App}. The effects of fitting for line of sight properties where isotropic dynamics is assumed were also considered, as well as the comparison of sensitivities of observational probes for anisotropies, from astrophysical systematics and from dark energy.

In addition to these, especially in the view of the latest Planck data,  the empirical evidence for the existence of cosmological anisotropies is strong. But, its physical origin is still unknown, although the most popular explanation being that the primordial fluctuations are deviating from isotropy \cite{1}. However, there is no clearly established physical mechanism that could lead to such deviations.

It is the purpose of the present paper to consider another possibility of explaining the presence of the anisotropies on a large cosmological scale, namely, the possibility that the two major components of the Universe, dark energy and dark matter, are not comoving. In standard cosmology the assumption that all constituents of the Universe move in the same frame, with the same four-velocity, is a fundamental (but not rigorously justifiable) assumption. This allows adopting a frame, which is comoving with matter in all its forms, in which the components of the four velocities $u^{\mu }$ of all components can be chosen as $u^{\mu }=(1,0,0,0)$. Consequently, the global thermodynamic parameters of the Universe are just the sum of the individual thermodynamic parameters. Thus, for example, in the standard model the energy density of the Universe $\rho $ is given by $\rho =\sum _{i=1}^{n}{\rho _i}$, where $\rho _i$, $i=1,2,..,n$ are the energy densities of the individual components.

However, there is no basic physical or observational principle that would require that all components in the Universe should have the same four-velocity, that is, one could describe the dynamics of the Universe in a single frame, comoving with all the constituent components. In this paper we will investigate the situation in which the Universe may be modeled as a mixture of two non-interacting perfect fluids, namely dark energy and dark matter, respectively, possessing different four-velocities. This configuration is formally equivalent to a single anisotropic fluid \cite{Le,Ol, He, Ha}. Therefore, if there is a slight difference between the four-velocities of the dark energy and dark matter, the Universe would acquire some anisotropic characteristics, and its geometry will deviate from the standard FLRW one.

The thermodynamic parameters (energy density and pressure) of the non-comoving two-fluid Universe are obtained by reducing the system to a single fluid, described by an  anisotropic energy-momentum tensor. The energy density of the single fluid is greater than the sum of the energy densities of the dark energy and dark matter, respectively, and it contains a correction due to anisotropy. As a cosmological application of the model we consider an effective induced  Bianchi type I geometry, generated by the non-comoving expansion of the dark energy and dark matter, respectively. The general properties of the system are discussed, and the expression of some observationally important parameters (expansion and shear) are obtained.  By using a simple perturbative approach, some simple cases of non-comoving dark anergy and dark matter cosmological models are constructed.

The possibility that there may exist different rest frames for dark matter and dark energy has been also investigated in \cite{Mar}, where it is shown that the existence of large scale bulk flows  could be signaling the presence of moving dark energy at the time when photons decoupled from matter. Bianchi I type cosmological models, the evolution of the homogeneous perturbations  up to second order, and the effects on the CMB quadrupole were also considered.

The possibilities of observationally testing the non-comoving dark energy-dark matter are briefly discussed. In particular we analyze the imprints of the anisotropy on the CMB spectrum, and on the luminosity distance. The expressions of the distribution of the temperature anisotropies and of the quadrupole $C_2$ are explicitly obtained in terms of the metric perturbations of the background flat geometry. This opens the possibility of observationally constraining the parameters of the model by using the recent Planck results \cite{1}, as well as the type Ia supernovae data \cite{acc}.

The present paper is organized as follows. The reduction of a two-fluid system to a single effective anisotropic fluid description is discussed in Section \ref{sect2}. The cosmological gravitational field equations of the model are obtained in Section \ref{sect3}, where some general properties of the non-comoving two fluid Universe are also discussed. The evolution of the cosmological perturbations due to the non-comoving motion of the dark energy and dark matter are considered in Section \ref{sectn}. The observational implications of our results are investigated in Section \ref{obs}. We discuss and conclude our results in Section \ref{sect4}. Throughout this paper, we use the natural system of units with $c=8\pi G=1$.

\section{Anisotropic fluid description of the non-comoving dark energy--dark matter cosmological models}\label{sect2}

We assume that the Universe consists of a mixture of two basic fluid components: dark energy, with energy density and pressure $\rho _{DE}$ and $p_{DE}$, and four-velocity $u_{DE}^{\mu }$, respectively; and dark matter, with thermodynamic parameters $\rho _{DM}$ and $p_{DM}$, respectively, and four-velocity $u_{DM}^{\mu }$. In standard cosmology it is assumed that the two cosmological fluids are comoving, which implies
\be
u_{DE}^{\mu }\equiv  u_{DM}^{\mu }\equiv u^{\mu }.
 \ee
This condition allows to choose a comoving frame in the study of the cosmological dynamics, where the components of the four velocity of all components can be reduced to the form $u^{\mu }=(1,0,0,0)$. However, in the present study we relax this condition, by assuming that, at least in some periods of the cosmological evolution, the two fundamental components of the Universe may have had different four-velocities,  so that
\be
u_{DE}^{\mu }\neq  u_{DM}^{\mu }.
\ee

 The physical and thermodynamical properties of the  two-component cosmological fluid are described by the energy-momentum tensor $T^{\mu \nu }$ of the Universe, given by the sum of each individual components as
\begin{eqnarray}
T^{\mu \nu }=\left( \rho _{DE}+p_{DE}\right) u^{\mu }_{DE}u^{\nu }_{DE}-p_{DE}g^{\mu \nu
}+\left( \rho _{DM}+p_{DM}\right) u^{\mu }_{DM}u^{\nu }_{DM}-p_{DM}g^{\mu \nu }.
\label{emtensor}
\end{eqnarray}
The four-velocities of the dark energy and dark matter are normalized according to $u^{\mu }_{DE}u_{DE\;\mu }=1$ and $%
u^{\mu }_{DM}u_{DM\;\mu }=1$, respectively. In the case when $u^{\mu }_{DE}= u^{\mu}_{DM}\equiv w^{\mu }$, the energy-momentum tensor of the two fluid system takes the form
\be
T^{\mu \nu }=\left( \rho _{DE}+p_{DE}+\rho _{DM}+p_{DM}\right) w^{\mu }w^{\nu }-\left(p_{DE}+p_{DM}\right)g^{\mu \nu}.
\ee
In this case the dark energy-dark matter system reduces to an isotropic  single fluid system. Hence, if dark matter and dark energy have the same four-velocity, the thermodynamic parameters of the  dark energy-dark matter two fluid system are obtained by simply adding the enthalpies and the pressures of the individual components. For such a physical system one can always adopt a comoving frame, in which the components of the four-velocity are $w^{\mu }=(1,0,0,0)$, and the components of the energy-momentum tensor of the two fluid system are given by $T_0^0= \left( \rho _{DE}+\rho _{DM}\right)\delta _0^0$, and $T_i^i=-\left(p_{DE}+p_{DM}\right)\delta _i^i$, $i=1,2,3$, no summation upon $i$.

In the following, we investigate a more general case, in which $u^{\mu }_{DE}\neq  u^{\mu}_{DM}$, that is, the four-velocities of the dark matter and dark energy are different. In this case it is impossible to introduce a comoving frame, in which the thermodynamic parameters are obtained as the sum of the thermodynamic parameters of the individual components. Moreover,  we assume that the energy density and the pressure of the dark energy satisfy the conditions $\rho _{DE}+p_{DE}\neq 0$, that is, the dark energy cannot be described by a ``pure'' cosmological constant.

The study of the cosmological models described by an energy-momentum tensor having the form given by Eq.~(\ref{emtensor}) can be significantly simplified if we cast it into the standard form of perfect anisotropic fluids. This can be done by means of the
transformations $u^{\mu }_{DE}\rightarrow u^{\ast \mu }_{DE}$ and $u^{\mu }_{DM}\rightarrow u^{\ast \mu }_{DM}$, respectively, so that \cite{Le,Ol, He}
\begin{widetext}
\begin{equation}\label{tel}
\left(
\begin{array}{c}
u^{\ast \mu }_{DE} \\
u^{\ast \mu }_{DM}
\end{array}
\right) =\left(
\begin{array}{cc}
\cos \alpha  & \sqrt{\frac{\rho _{DM}+p_{DM}}{\rho _{DE}+p_{DE}}}\sin \alpha  \\
-\sqrt{\frac{\rho _{DE}+p_{DE}}{\rho _{DM}+p_{DM}}}\sin \alpha  & \cos \alpha
\end{array}
\right) \times \left(
\begin{array}{c}
u^{\mu }_{DE} \\
u^{\mu }_{DM}
\end{array}
\right) ,
\end{equation}
\end{widetext}
represents a ``rotation'' of the four-vector velocities in the $\left(
u^{\mu }_{DE},u^{\mu }_{DM} \right)$ velocity space. Explicitly, the transformations (\ref{tel}) take the form
\be
u^{\mu }_{DE}\rightarrow u^{\ast \mu }_{DE}=u^{\mu }_{DE}\cos \alpha +\sqrt{\frac{\rho
_{DM}+p_{DM}}{\rho _{DE}+p_{DE}}}u^{\mu }_{DM}\sin \alpha ,  \label{te1}
\end{equation}
\begin{equation}
u^{\mu }_{DM}\rightarrow u^{\ast \mu }_{DM}=u^{\mu }_{DM}\cos \alpha -\sqrt{\frac{\rho
_{DE}+p_{DE}}{\rho _{DM}+p_{DM}}}u^{\mu }_{DE}\sin \alpha \,,  \label{te2}
\end{equation}
respectively.

The velocity transformations given by Eq.~(\ref{tel})  leave the quadratic form $\left(
\rho _{DE}+p_{DE}\right) u^{\mu }_{DE}u^{\nu }_{DE}+\left( \rho _{DM}+p_{DM}\right) u^{\mu
}_{DM}u^{\nu }_{DM}$ invariant. Thus,
\be
T^{\mu \nu }\left( u^{\mu}_{DE},u^{\nu }_{DM}\right) =T^{\mu \nu
}\left( u^{\ast \mu}_{DE},u^{\ast \nu }_{DM}\right) .
\ee
Next we choose
 $u^{\ast \mu }_{DE}$ and $u^{\ast \mu }_{DM}$ such
that one becomes timelike, while the other is spacelike. Moreover, we assume that the two transformed four-vector velocities satisfy the condition
\be\label{cond}
u^{\ast \mu }_{DE}u_{DM\;\mu }^{\ast }=0.
\ee
With the use of Eqs.~(\ref{te1})-(\ref{te2}) and of the condition given by Eq.~(\ref{cond}), we obtain first the rotation angle as
\begin{equation}\label{alpha}
\tan 2\alpha =2\frac{\sqrt{\left( \rho _{DE}+p_{DE}\right) \left( \rho
_{DM}+p_{DM}\right) }}{\rho _{DE}+p_{DE}-\left( \rho _{DM}+p_{DM}\right) }u^{\mu
}_{DE}u_{DM\;\mu }.
\end{equation}
 If the angle $\alpha $ does not have the specific form given by Eq.~(\ref{alpha}), one cannot find a spacelike  $u^{\ast \mu }_{DE}$, and a timelike $u^{\ast \mu }_{DM}$ transformed dark energy and dark matter velocity, respectively. Note that if $\rho _{DE}+p_{DE}=0$, which corresponds to the case of the cosmological constant, the rotation angle is $\alpha =0$, and therefore the rotation in the velocity space reduces to the identical transformation. In the case of a pure cosmological constant, since $\rho _{DE}+p_{DE}=0$, the term $\left(\rho _{DE}+p_{DE}\right)u^{\mu }_{DE}u^{\nu }_{DE}$ in the energy-momentum tensor is identically equal to zero, $\left(\rho _{DE}+p_{DE}\right)u^{\mu }_{DE}u^{\nu }_{DE}\equiv 0$, and the four-velocity of the cosmological constant does not appear in the formalism. Therefore the effective energy -momentum tensor of system consisting of a pure cosmological constant plus dark matter reduces to the energy-momentum tensor of a single isotropic fluid.

Now, by defining the quantities
\begin{eqnarray}
V^{\mu }&=&\frac{u^{\ast \mu }_{DE}}{\sqrt{u^{\ast \alpha }_{DE}u_{DE\;\alpha }^{\ast }}},
   \\
\chi ^{\mu }&=&\frac{u^{\ast \mu }_{DM}}{\sqrt{-u^{\ast \alpha }_{DM}u_{DM\;\alpha }^{\ast }}},
\end{eqnarray}
and
\begin{eqnarray}
\varepsilon =T^{\mu \nu }V_{\mu }V_{\nu }=\left( \rho _{DE}+p_{DE}\right)
u^{\ast \alpha }_{DE}u_{DE\;\alpha }^{\ast }-\left( p_{DE}+p_{DM}\right),
\end{eqnarray}
\begin{eqnarray}
\Psi =T^{\mu \nu }\chi _{\mu }\chi _{\nu }
   =\left( p_{DE}+p_{DM}\right)-\left( \rho_{DM}+p_{DM}\right) u^{\ast \alpha }_{DM}u_{DM\;\alpha }^{\ast },
\end{eqnarray}
\be
\Pi =p_{DE}+p_{DM},
\ee
respectively,  where $\varepsilon $ is the energy density and $\Psi $ the radial pressure of the anisotropic Universe, then the energy-momentum tensor of the two non-interacting perfect fluids
can be written as
\begin{equation}
T^{\mu \nu }=\left( \varepsilon +\Pi \right) V^{\mu }V^{\nu }-\Pi g^{\mu \nu
}+\left( \Psi -\Pi \right) \chi ^{\mu }\chi ^{\nu },  \label{tens}
\end{equation}
where $V^{\mu }V_{\mu }=1=-\chi ^{\mu }\chi _{\mu }$ and $\chi ^{\mu }V_{\mu }=0$ \cite{Le,Ol, He}.
The energy-momentum tensor of the two-fluid cosmological model, in which the components have different four-velocities, given by Eq.~(\ref{tens}), is the standard form for anisotropic fluids \cite{He}.

The energy density $\varepsilon $ and the radial pressure $\Psi $ of the anisotropic Universe are
given by
\begin{widetext}
\bea\label{d1}
\varepsilon & =& \frac{1}{2}\left( \rho _{DE}+\rho _{DM}-p_{DE}-p_{DM}\right) +\nonumber\\
&&\frac{%
1}{2}\sqrt{\left( \rho _{DE}+p_{DE}+\rho _{DM}+p_{DM}\right) ^{2}+4\left( \rho
_{DE}+p_{DE}\right) \left( \rho _{DM}+p_{DM}\right) \left[ \left( u^{\mu }_{DE}u_{DM\;\mu
}\right) ^{2}-1\right] },  \label{eps1}
\eea
and
\bea\label{psin}
\Psi &=&-\frac{1}{2}\left( \rho _{DE}+\rho _{DM}-p_{DE}-p_{DM}\right) +\nonumber\\
&&\frac{1}{2%
}\sqrt{\left( \rho _{DE}+p_{DE}+\rho _{DM}+p_{DM}\right) ^{2}+4\left( \rho
_{DE}+p_{DE}\right) \left( \rho _{DM}+p_{DM}\right) \left[\left( u^{\mu }_{DE}u_{DM\;\mu
}\right) ^{2}-1\right]},  \label{sig}
\eea
\end{widetext}
respectively \cite{Le,Ol, He}. The energy density of the anisotropic effective fluid, given by Eq.~(\ref{d1}) explicitly depends on the four-velocities of the two fluids. Hence, the energy density of the rotated fluid also depends on the kinetic energy of the two fluids, and this dependence makes the rotated energy different from the sum of the rest mass energies of the component fluids. When $ u^{\mu }_{DE}=u_{DM}^\mu$, then $u^{\mu }_{DE}u_{DM\;\mu
}=1$, and the expressions of the effective anisotropic energy and pressure reduce to the sum of the energy densities of the dark anergy and of the dark matter, respectively, $\varepsilon =\rho _{DE}+\rho _{DM}$, $\Psi =p_{DE}+p_{DM}$.

Since the two cosmological fluids, dark energy and dark matter, have different four-velocities, we may write
\be
u^{\mu }_{DE}u_{DM\;\mu }=1+\frac{b}{2},
\ee
where generally $b$ is an arbitrary function of the radial and time coordinates. The functional form of $b$ can be obtained from Eq.~(\ref{alpha}), which gives first
\bea
u^{\mu }_{DE}u_{DM\;\mu }=\frac{1}{2}\tan 2\alpha \Bigg[ \sqrt{\frac{\rho _{DE}+p_{DE}}{
\rho _{DM}+p_{DM}}}-\sqrt{\frac{\rho _{DM}+p_{DM}}{\rho _{DE}+p_{DE}}}\Bigg],
\eea
while $b$ can be obtained as
\begin{equation}
b=\left[ \sqrt{\frac{\rho _{DE}+p_{DE}}{%
\rho _{DM}+p_{DM}}}-\sqrt{\frac{\rho _{DM}+p_{DM}}{\rho _{DE}+p_{DE}}}\right]\tan 2\alpha -2\texttt{}.
\end{equation}

In terms of $b$ Eqs.~(\ref{d1}) and (\ref{psin}) can be written as
\bea\label{epsn}
\varepsilon & =& \frac{1}{2}\left( \rho _{DE}+\rho _{DM}-p_{DE}-p_{DM}\right) +
\frac{%
1}{2}\left( \rho _{DE}+p_{DE}+\rho _{DM}+p_{DM}\right) \sqrt{1+4b\left(1+\frac{b}{4}\right)\frac{\left( \rho
_{DE}+p_{DE}\right) \left( \rho _{DM}+p_{DM}\right)}{\left( \rho _{DE}+p_{DE}+\rho _{DM}+p_{DM}\right) ^{2}}  },  \nonumber\\
\eea
and
\bea\label{psin}
\Psi &=&-\frac{1}{2}\left( \rho _{DE}+\rho _{DM}-p_{DE}-p_{DM}\right) +
\frac{1}{2%
}\left( \rho _{DE}+p_{DE}+\rho _{DM}+p_{DM}\right)
\sqrt{1+4b\left(1+\frac{b}{4}\right)\frac{\left( \rho
_{DE}+p_{DE}\right) \left( \rho _{DM}+p_{DM}\right)}{\left( \rho _{DE}+p_{DE}+\rho _{DM}+p_{DM}\right) ^{2}} }, \nonumber\\
\eea
respectively.

By assuming that the physical parameters of the dark energy and dark matter satisfy the
condition
\begin{equation}
4b\left( 1+\frac{b}{4}\right) \frac{\left( \rho _{DE}+p_{DE}\right) \left(
\rho _{DM}+p_{DM}\right) }{\left( \rho _{DE}+\rho _{DM}+p_{DE}+p_{DM}\right) ^{2}}%
\ll 1,
\end{equation}
by series expanding the square root in Eqs.~(\ref{epsn}) and (\ref{psin}), in the first order of approximation,  the energy density, the radial and the tangential pressures of the Universe can be obtained as
as
\begin{equation}\label{eps}
\varepsilon =\rho _{DE}+\rho _{DM}+2b\left( 1+\frac{b}{4}\right)
\frac{\left( \rho _{DE}+p_{DE}\right) \left( \rho _{DM}+p_{DM}\right) }{\left(
\rho _{DE}+\rho _{DM}+p_{DE}+p_{DM}\right) },
\end{equation}
\begin{equation}
\Psi =p_{DE}+p_{DM}+2b\left( 1+\frac{b}{4}\right) \frac{\left( \rho
_{DE}+p_{DE}\right) \left( \rho _{DM}+p_{DM}\right) }{\left( \rho _{DE}+\rho
_{DM}+p_{DE}+p_{DM}\right) },
\end{equation}
and
\begin{equation}
\Pi =p_{DE}+p_{DM},
\end{equation}
respectively. As one can readily verify from Eq.~(\ref{eps}), the energy density of the non-comoving two fluid Universe is different from the sum of the energy densities of the component fluids.

\section{Anisotropic two fluid cosmological dynamics}\label{sect3}

In the present Section, we consider some cosmological applications of the non-comoving dark energy-dark matter cosmological model. As a first step in our analysis, we write down the gravitational field equations corresponding to the anisotropic evolution of the Universe. The general properties of the model are analyzed, and it is shown that the non-comoving nature of the cosmological dynamics affects the observational parameters of the Universe, by inducing some specific anisotropic effects in the evolution of both expansion and shear parameters. Finally, an explicit dark energy model is considered, and the general solution of the field equations, describing the dark energy-dark matter mixture is obtained.

\subsection{Gravitational field equations of the two non-comoving fluid cosmological model}

For the two-fluid energy-momentum tensor, with components given by Eqs.~(\ref{tens}), the Einstein gravitational field equations can be written as \cite{Ol}
\be
R_{\mu \nu}V^{\mu }V^{\nu }=\frac{1}{2}\left(\varepsilon +2\Pi +\psi\right),
\ee
\be
R_{\mu \nu}V^{\mu }h^{\nu }_{\sigma }=0,
\ee
\bea
R_{\mu \nu}h^{\mu }_{\sigma }h^{\nu }_{\lambda }&=&\frac{1}{2}\left[\varepsilon -\frac{1}{3}\left(2\Pi +\Psi\right)\right]h_{\sigma \lambda }
+\left(\Psi -\Pi\right)\left(\chi _{\sigma}\chi _{\lambda }-\frac{1}{3}h_{\sigma \lambda }\right),
\eea
where $h^{\mu \nu}=g^{\mu \nu}-V^{\mu }V^{\nu }$. The conservation of the energy-momentum tensor $\nabla _{\mu }T_{\nu}^{\mu }=0$ yields
\be
\dot{\varepsilon}+\left(\varepsilon +\Pi\right)\nabla _{\mu }V^{\mu}-\left(\Psi-\Pi\right)\chi ^{\prime \mu }V_{\mu }=0,
\ee
\be
\left(\varepsilon +\Pi \right)\dot{V}^{\mu}+h^{\mu \nu}\nabla _{\nu }\Pi +\left(\Psi -\Pi\right)^{\prime}\chi ^{\mu }
  +\left(\Psi -\Pi\right)\left(\nabla _{\nu}\chi ^{\nu }\chi ^{\mu }+\chi _{\nu }^{\prime}h^{\nu \mu }\right)=0,
\ee
where the overdot and the prime are defined as $\dot{\left(\right)}=\nabla _{\mu }\left(\right)V^{\mu }$, and $\left(\right)^{\prime}=\nabla _{\mu }\left(\right)\chi ^{\mu }$, respectively.

In the following, we assume that the Universe is homogeneous on a large scale, and therefore all the cosmological parameters (energy densities, pressures, and four-velocities) are functions of the cosmological time only. The assumption of homogeneity allows the reduction of the auxiliary quantities $V$ and $\chi $ to a form that is equivalent to the choice of a frame ``comoving'' with them. In comoving Cartesian coordinates $x^{0}=t$, $x^{1}=x$, $%
x^{2}=y $, and $x^{3}=z $ we may choose $V^{1}=V^{2}=V^{3}=0$, $%
V^{0}V_{0}=1$, and $\chi ^{0}=\chi ^{1}=\chi ^{2}=0$, $\chi ^{3}\chi _{3}=-1$, respectively
\cite{Le, Ol, He}.  Therefore the components of the energy-momentum of two non-comoving cosmological
perfect fluids take the form
\begin{equation}\label{tenss}
T_{0}^{0}=\varepsilon, \qquad  T_{1}^{1}=T_2^2=-\Pi , \qquad T_{3}^{3}=-\Psi ,
\end{equation}
where $\varepsilon $ is the total energy-density of the
mixture of fluids, $\Psi =P_{z}$ is the pressure along the $z$
direction, while $\Pi =P_{x }=P_y$ is the  pressure along the  $x$ and $y$
directions. Since the two-fluid dark energy-dark matter cosmological mixture represents an anisotropic
fluid, on a cosmological scale the corresponding geometry is also
anisotropic.  The energy-momentum tensor components given by Eqs.~(\ref{tenss}) shows that the pressure $\Psi$ along the $z-$direction may be different with respect to the pressure $\Pi$ on the $x-y$ plane.

In a homogeneous Universe with a  two-fluid mixture the simplest geometry exhibiting this symmetry of the energy-momentum tensor is the flat Bianchi type I geometry, with the line element given by
\begin{equation}
ds^{2}=dt^{2}-a_{1}^{2}(t)dx^{2}-a_{2}^{2}(t)dy^{2}-a_{3}^{2}(t)dz^{2},
\label{7}
\end{equation}%
where $a_i(t)$ ($i=1,2,3$) are the directional scale factors. In this geometry the gravitational field equations take the form
\begin{equation}
3\dot{H}+H_{1}^{2}+H_{2}^{2}+H_{3}^{2}=-\frac{1}{2}\left( \varepsilon +\Psi
+2\Pi \right) ,  \label{8}
\end{equation}%
\be
\frac{1}{V}\frac{d}{dt}\left( VH_{i}\right)  =\frac{1}{2}\left(
\varepsilon -\Pi \right) , \qquad i=1,2 , \label{9}
\ee
and
\be\label{101}
\frac{1}{V}\frac{d}{dt}%
\left( VH_{3}\right) =\frac{1}{2}\left( \varepsilon -\Psi \right) ,
\ee
respectively. We have denoted
\begin{equation}
V=a_{1}a_{2}a_{3}, \qquad H_{i}=\frac{\dot{a}_{i}}{a_{i}}, \qquad i=1,2,3
\end{equation}
and
\begin{equation}
H=\frac{1}{3} \left(\sum_{i=1}^{3}H_{i}\right)=\frac{\dot{V}}{3V}
\end{equation}
respectively. The expansion parameter $\theta $ is given by $\theta =3H$. From the conservation of the total energy-momentum tensor of the two-fluid cosmological model we obtain the evolution equation of the energy density of the system, given by
\be
\dot{\varepsilon}+3\left(\varepsilon +\Pi\right)H-\left(\Psi -\Pi \right)H_3=0.
\ee

\subsection{General properties of the anisotropic Bianchi I two fluid cosmological model}

From Eqs.~(\ref{9}) it follows that for the considered form of the anisotropic energy-momentum tensor one can take $a_1(t)=a_2(t)$ without any loss of generality. Therefore the expansion parameter of the Universe can be obtained as
\be\label{th}
\theta =2\frac{\dot{a}_1}{a_1}+\frac{\dot{a}_3}{a_3},
\ee
while the shear scalar $\sigma $ of the Universe is given by
\be\label{sh}
\sigma =\frac{1}{\sqrt{3}}\left(\frac{\dot{a}_3}{a_3}-\frac{\dot{a}_1}{a_1}\right).
\ee
Equations~(\ref{th}) and (\ref{sh}) allow the determination of the directional Hubble parameters $H_1$ and $H_3$ in terms of observable cosmological parameters as \be
H_1=\frac{\theta }{3}-\frac{1}{\sqrt{3}}\sigma,\qquad  H_3=\frac{\theta }{3}+\frac{2}{\sqrt{3}}\sigma,
\ee
respectively.

The large-time behavior of an anisotropic cosmological model can be obtained from the study of the  anisotropy
parameter $A$, defined by
\begin{equation}
A=\frac{1}{3}\sum_{i=1}^{3}\left( \frac{H_{i}-H}{H}\right) ^{2}.
\end{equation}
If $A=0$ the cosmological model is isotropic.
The anisotropic cosmological pressure differences can be obtained, in terms of the observable quantities, as
\be
\Psi -\Pi =\sqrt{3}\sigma +\theta \sigma,
\ee
or, equivalently,
\begin{widetext}
\be
\sqrt{3}\sigma +\theta \sigma=\frac{1}{2}\Bigg\{-\left(\rho _{DE}+\rho _{DM}\right)+
\sqrt{\left( \rho _{DE}+p_{DE}+\rho _{DM}+p_{DM}\right) ^{2}+4\left( \rho
_{DE}+p_{DE}\right) \left( \rho _{DM}+p_{DM}\right) \left[\left( u^{\mu }_{DE}u_{DM\;\mu
}\right) ^{2}-1\right]}\Bigg\}.
\ee
\end{widetext}

In order to describe the deviations from a comoving motion, we introduce  the parameter $\beta (t)$, defined as
\be
\beta (t)\equiv \frac{\left( \rho
_{DE}+p_{DE}\right) \left( \rho _{DM}+p_{DM}\right) \left[\left( u^{\mu }_{DE}u_{DM\;\mu
}\right) ^{2}-1\right]}{\left( \rho _{DE}+p_{DE}+\rho _{DM}+p_{DM}\right) ^{2}}.
\ee

By assuming that $\beta (t)$,  is small, $\beta (t)\ll 1$, to first order in $\beta$, we obtain the following expressions of the energy density and $z$-direction pressure of the two non-comoving fluid cosmological model,
\be
\varepsilon\approx \rho _{DE}+\rho _{DM}+\beta \left(\rho _{DE}+p_{DE}+\rho _{DM}+p_{DM}\right),
\ee
and
\be
\Psi=p_{DE}+p_{DM}+\beta \left(\rho _{DE}+p_{DE}+\rho _{DM}+p_{DM}\right),
\ee
respectively. Therefore the total  energy density and pressure of the two-fluid cosmological model is not equal to the sum of the energy densities and pressures of the individual matter components of the system.

By adding Eqs.~(\ref{9}) and (\ref{101})  we obtain
\begin{equation}
\dot{H}+3H^{2}=\frac{1}{2}\varepsilon -\frac{1}{3}\left(\Pi+\frac{\Psi }{2}\right),  \label{10}
\end{equation}%
or, equivalently,
\be\label{41}
\ddot{V}+\left(\Pi+\frac{1}{2}\Psi -\frac{3}{2}\varepsilon \right)V=0.
\ee
These expression will be considered in the next section.

\section{Perturbative approach to non-comoving two fluid cosmological models}\label{sectn}

In the following, we consider a simple perturbative solution of the field equations (\ref{8})-(\ref{101}) by assuming that the anisotropic expansion due to the non-comoving motion of dark energy and dark matter represents a small perturbation of the background isotropic, flat FLRW geometry, described by the isotropic scale factor $a(t)$. We also assume that the expansion rates along the $x$ and $y$ axes are the same, so that  $a_1(t)=a_2(t)$, and the perturbations along these axes are identical. Therefore the scale factor of the perturbed Bianchi type I geometry can be given as
\be
a_i(t)=a(t)+\delta _i(t), \qquad i=1,2,3,
 \ee
 where $\delta _i(t)$, $i=1,2,3$, are small corrections in the scale factor $a(t)$ of the isotropic Universe, induced by the small differences between the four-velocities of the dark matter and dark energy. Moreover, we assume that
 \be
 \delta _1(t)=\delta _2(t).
 \ee
The Hubble parameter of the background isotropic space is $H_0(t)=\dot{a}/a$.  For the directional Hubble
parameters of the anisotropic expansion we obtain
 \bea\label{Hi}
 H_i(t)=\frac{\dot{a}(t)+\dot{\delta}_i(t)}{a(t)+\delta _i(t)}\approx H_0(t)\left[1-\frac{\delta _i(t)}{a(t)}\right]+\frac{\dot{\delta _i}(t)}{a(t)},
  \eea
where $i=1,2,3$. Thus, the mean Hubble factor can be obtained as
  \be
  H(t)=\frac{1}{3}\left[3H_0(t)-H_0(t)\frac{\delta (t)}{a(t)}+\frac{\dot{\delta }(t)}{a(t)}\right],
  \ee
 where we have denoted
 \be
 \delta (t)=2\delta _1(t)+\delta _3(t),
 \ee
while the square of the directional Hubble factors  is given by
\be
H_i^2(t)\approx H_0^2(t)-2H_0^2(t)\frac{\delta _i(t)}{a(t)}+2H_0(t)\frac{\dot{\delta}_i(t)}{a(t)},
 \ee
where $i=1,2,3$.

For the volume $V(t)$ of the Universe we obtain
 \be
V(t)=a^3(t)\left[1+\frac{\delta (t)}{a(t)}\right],
 \ee
and $1/V(t)\approx 1/a^3(t)$.

Hence, to a first order approximation, the gravitational field equations describing the slightly perturbed isotropic flat FLRW Universe become
\bea
3\dot{H}_0(t)+3H_0^2(t)-\left[\dot{H}_0(t)+H_0^2(t)\right]\frac{\delta (t)}{a(t)}+\frac{\ddot{\delta}(t)}{a(t)}&=&
-\frac{1}{2}\Big[\rho _{DE}+\rho _{DM}+
3\left(p_{DE}+p_{DM}\right) \nonumber\\
&&+2\beta  \left(\rho _{DE}+p_{DE}+\rho _{DM}+p_{DM}\right)\Big],
\eea
\bea
\frac{1}{a^3(t)}\frac{d}{dt}\left[a^3(t)H_0(t)\right]+\frac{1}{a^3(t)}\frac{d}{dt}\left[a^2(t)H_0(t)\delta _{+}(t)\right]
+\frac{1}{a^3}\frac{d}{dt}\left[a^2(t)\dot{\delta }_1(t)\right]
   \nonumber\\
=\frac{1}{2}\Big[\rho _{DE}-p_{DE}+\rho _{DM}-p_{DM}
+\beta \left(\rho _{DE}+p_{DE}+\rho _{DM}+p_{DM}\right)\Big],
\eea
and
\bea
\frac{1}{a^3(t)}\frac{d}{dt}\left[a^3(t)H_0(t)\right]+\frac{2}{a^3}\frac{d}{dt}\left[a^2(t)H_0(t)\delta_{1}(t)\right]
  +\frac{1}{a^3}\frac{d}{dt}\left[a^2(t)\dot{\delta}_{3}(t)\right]=
 \frac{1}{2}\left(\rho _{DE}-p_{DE}+\rho _{DM}-p_{DM}\right),
\eea
respectively, where we have denoted $\delta _{+}(t)=\delta _1(t)+\delta _2(t)$.
Equation (\ref{41}), giving the time evolution of the volume of the Universe, takes the form
\bea
\frac{d^2}{dt^2}a^3(t)+\frac{3}{2}\left[a^3(t)\left(p_{DE}-\rho _{DE}+p_{DM}-\rho_{DM}\right)\right]+
\frac{d^2}{dt^2}a^2(t)\delta (t)
+\frac{3}{2}\left[a^2(t)\left(p_{DE}-\rho _{DE}+p_{DM}-\rho_{DM}\right)\right]
\delta (t)
     \nonumber\\
- \frac{1}{2}
 \left[a^3(t)\beta \left(\rho _{DE}+p_{DE}+\rho _{DM}+p_{DM}\right)\right]=0.
\eea

The gravitational field equations of the background, isotropic geometry, are the standard Friedmann equations, determining the scale factor $a$ as
\be\label{fr1}
3\frac{\dot{a}^2(t)}{a^2(t)}=\rho _{DE}+\rho _{DM},
\ee
\be\label{fr2}
2\frac{\ddot{a}(t)}{a(t)}+\frac{\dot{a}^2(t)}{a(t)}=-p_{DE}-p_{DM},
\ee
respectively.
With the use of  Eqs.~(\ref{fr1}) and (\ref{fr2}) we immediately obtain
\be
\frac{1}{a^3(t)}\frac{d\left[a^3(t)H_0(t)\right]}{dt}=\frac{1}{2}\left(\rho _{DE}-p_{DE}+\rho _{DM}-p_{DM}\right),
 \ee
 \be
 3\dot{H}_0(t)+3H_0^2(t)=-\frac{1}{2}\left[\rho _{DE}+\rho _{DM}+3\left(p_{DE}+p_{DM}\right)\right],
  \ee
  and
 \be
  \frac{d^2a^3(t)}{dt^2}+\frac{3}{2}\left[a^3(t)\left(p_{DE}-\rho _{DE}+p_{DM}-\rho_{DM}\right)\right]=0,
   \ee
respectively.

Therefore, the gravitational field equations describing the time evolution of a cosmological model with a
small anisotropy in  the $z$-direction, induced by the non-comoving nature of dark energy and dark matter,
take the form
\bea\label{pert1}
\frac{\ddot{\delta}(t)}{a(t)}-\frac{\ddot{a}(t)}{a(t)}\frac{\delta (t)}{a(t)}=
-\beta \left(\rho _{DE}+p_{DE}+\rho _{DM}+p_{DM}\right),
\eea
\bea\label{pert2}
\frac{1}{a^3(t)}\frac{d}{dt}\left[a^2(t)H_0(t)\delta _{+}(t)\right]+\frac{1}{a^3}\frac{d}{dt}\left[a^2(t)\dot{\delta }_1(t)\right]=
\frac{1}{2}\beta \left(\rho _{DE}+p_{DE}+\rho _{DM}+p_{DM}\right),
\eea
\be\label{pert3}
\frac{2}{a^3}\frac{d}{dt}\left[a^2(t)H_0(t)\delta_{1}(t)\right]+\frac{1}{a^3}\frac{d}{dt}\left[a^2(t)\dot{\delta}_{3}(t)\right]=0.
\ee
Equations (\ref{pert1})-(\ref{pert3}) represent a system of three differential equations for the three unknown functions $\left(\delta _1(t),\delta _3(t),\beta (t)\right)$.

It will be useful to introduce the deceleration parameter $q$ of the non-comoving two fluid anisotropic Universe, which is defined as
\be
q=\frac{d}{dt}\frac{1}{H(t)}-1,
\ee
and for the case of the small deviation from isotropy takes the form
\be
q=q_0+\frac{1}{3}\frac{d}{dt}\frac{1}{aH_0}\left[\delta (t)-\frac{\dot{\delta }(t)}{H_0}\right],
\ee
where $q_0=d\left[1/H_0(t)\right]/dt-1=-a(t)\ddot{a}(t)/\dot{a}^2(t)$ is the deceleration parameter of the isotropic FLRW Universe.

Now, Eq.~(\ref{pert3}) can be immediately integrated to give
\be
2H_0(t)\delta _1(t)+\dot{\delta }_3(t)=\frac{C}{a^2(t)},
\ee
where $C$ is an arbitrary constant of integration. In order to determine the value of $C$ we assume that at $t=t_0$ the Universe was in an approximate isotropic state, so that $H_i\left(t_0\right)\approx H_0\left(t_0\right)$, $i=1,2,3$, with a small anisotropic perturbation  $\delta _i \left(t_0\right)=\delta _{0i}$, $i=1,2,3$. Then from Eq.~(\ref{Hi}) we have $\dot{\delta }_i\left(t_0\right)=H_0\left(t_0\right)\delta _{0i}$, $i=1,2,3$. For the initial values of $\delta $ we obtain $\delta \left(t_0\right)=2\delta _1\left(t_0\right)+\delta _3\left(t_0\right)=\delta _0$,  while $\dot{\delta }\left(t_0\right)=H_0\left(t_0\right)\delta _0$. Hence the value of the integration constant $C$ is given by
\bea
C=a^2\left(t_0\right)\left[2H_0\left(t_0\right)\delta _{01}+H_0\left(t_0\right)\delta _{03}\right]
  =a^2\left(t_0\right)H_0\left(t_0\right)\delta _0.
\eea

In order to obtain the evolution of the perturbation of the total perturbation due to the non-comoving motion of the dark energy and dark matter, we multiply Eq.~(\ref{pert2}) by two, and add the resulting equation  to Eq.~(\ref{pert3}), from which we obtain
\bea\label{fin2}
\frac{2}{a^3(t)}\frac{d}{dt}\left[a^2(t)H_0(t)\delta (t)\right]+\frac{1}{a^3}\frac{d}{dt}\left[a^2(t)\dot{\delta }(t)\right]=
  \beta \left(\rho _{DE}+p_{DE}+\rho _{DM}+p_{DM}\right).
\eea

Then, by eliminating the term containing the energy densities between Eqs.~(\ref{pert1}) and (\ref{fin2}), we obtain the second order differential equation  governing the behavior of the total perturbation $\delta $ due to the
non-comoving motion of the dark energy and dark matter as
\be\label{eqfin}
2\ddot{\delta }(t)+4H_0(t)\dot{\delta }(t)+\left[\dot{H}_0(t)+3H_0^2(t)\right]\delta (t)=0.
\ee
Equation~(\ref{eqfin}) must be integrated together with the initial conditions
\be
\delta \left(t_0\right)=\delta _0, \qquad \dot{\delta }\left(t_0\right)=H_0\left(t_0\right)\delta _0.
\ee

The behavior of the perturbations to isotropy induced by the non-comoving flow of the dark energy and dark matter is determined by the Hubble parameter of the isotropic FLRW background. Once the function $\delta (t)$ is known, the time evolution of the perturbations along the  $z-$axis is determined by the first order linear differential equation
\be
\dot{\delta }_3(t)-H_0\delta _3(t)=\frac{C}{a^2(t)}-H_0\delta (t),
\ee
with the general solution given by
\be
\delta _3(t)=a(t)\left\{C_1+\int{\left[\frac{C}{a^3(t)}-\frac{H_0(t)}{a(t)}\delta (t)\right]dt}\right\},
\ee
where $C_1$ is an arbitrary integration constant to be determined from the initial condition $\delta _3\left(t_0\right)=\delta _{03}$. The perturbation from  isotropy along the $x$ and $y$ axes can be obtained from
\be\label{73}
\delta _1(t)=\frac{1}{2}\left[\delta (t)-\delta _3(t)\right],
\ee
while the anisotropy parameter $\beta $ can be obtained as
\be
\beta (t)=\frac{2H_0(t)\dot{\delta }(t)+\left[3\dot{H}_0(t)/2+5H_0^2(t)/2\right]\delta (t)}{a\left(\rho _{DE}+p_{DE}+\rho _{DM}+p_{DM}\right)}.
\ee
Hence we have obtained the general solution of the perturbed gravitational field equations due to the non-comoving motion of the dark energy and dark matter.

As simple applications of the present formalism we will consider two cases, namely, the case of the constant $H_0$, corresponding to an exponentially expanding Universe, $a(t)=\exp\left(H_0t\right)$, and of the power law expanding isotropic FLRW Universe, with $a(t)=t^n$. Note, however, that in order to obtain the expression of the anisotropy parameter $\beta $, the energy densities and pressures of the dark energy and dark matter must also be known.

\subsection{Specific solution I: $a(t)=\exp\left(H_0t\right)$}

If $H_0={\rm constant}$, then the general solution of Eq.~(\ref{eqfin}) is given by
\be
\delta (t)=e^{-H_0t}\left[C_2\cos\left(\sqrt{8}H_0t\right)+C_3\sin\left(\sqrt{8}H_0t\right)\right],
\ee
where $C_2$ and $C_3$ are arbitrary constants of integration to be determined from the initial conditions $\delta \left(t_0\right)=\delta _0$, and $\dot{\delta }\left(t_0\right)=H_0\left(t_0\right)\delta _0$, respectively. For an exponentially expanding background Universe the magnitude of the total anisotropy $\delta $ decreases exponentially in time.

By using these expressions of $\delta (t)$ the perturbations $\delta _1 (t)$ and $\delta _3(t)$ along the $x,y$ and $z$ axes can be explicitly computed for these particular cosmological models. For the de Sitter-type background expansion, we obtain
\begin{widetext}
\be
\delta _3(t)=C_1e^{H_0t}+\frac{e^{-2 H_0 t} \left\{H_0 e^{H_0 t}
   \left[\left(C_3-\sqrt{2} C_2\right) \sin \left(2 \sqrt{2} H_0
   t\right)+\left(C_2+\sqrt{2} C_3\right) \cos \left(2 \sqrt{2} H_0
   t\right)\right]-2 C\right\}}{6 H_0}\,,
   \ee
and
\be
\delta _1(t)=\frac{e^{-2 H_0 t} \left\{2 \left(C-3 C_1 H_0 e^{3 H_0t}\right)+H_0e^{H_0t} \left[\left( 5C_2-\sqrt{2} C_3\right) \sin \left(2
   \sqrt{2} H_0 t\right)+\left(5C_2+\sqrt{2} C_3\right) \cos \left(2 \sqrt{2} H_0 t\right)\right]\right\}}{12 H_0} ,
   \ee
   \end{widetext}
respectively.
The time evolution of the anisotropic perturbation along the $z-$axis is proportional to the scale factor of the isotropic background, and thus for a de Sitter-type Universe the anisotropy in the $z$ direction will increase in time, making the Universe more and more anisotropic. On the other hand, since in the large time limit $\delta $ tends to zero,  $\lim_{t\rightarrow \infty}\delta =0$,  from Eq.~(\ref{73}) it follows that $\lim_{t\rightarrow \infty}{\delta _1(t)}=-(1/2)\lim_{t\rightarrow \infty}{\delta _3(t)}=-(1/2)\lim_{t\rightarrow \infty}{C_1\exp\left(H_0t\right)}$. However, in these large time limits the formalism developed in the present paper may not be applicable, since all the presented results have been derived under the assumption $\delta _i \ll a$, $i=1,2,3$.

\subsection{Specific solution II: power law expansion}

For a power law expansion, with the scale factor of the form $a(t)=t^n$, Eq.~(\ref{eqfin}) becomes
\be
2\ddot{\delta}+\frac{4n}{t}\dot{\delta}+\frac{n\left(3n-1\right)}{t^2}\delta (t)=0,
\ee
with the general solution given by
\be
\delta (t)=C_4t^{r_1}+C_5t^{r_2},
\ee
where
\be
r_{1,2}=\frac{1-2n\pm \sqrt{2}\sqrt{2-n-n^2}}{2}.
\ee

In the case of the power law background expansion we obtain
   \be
   \delta _3(t)=C_1 t^n+t \left(\frac{c t^{-2 n}}{1-3 n}+\frac{C_4H_0
   t^{r_1}}{n-r_1-1}+\frac{C_5H_0
   t^{r_2}}{n-r_2-1}\right),
   \ee
while for $\delta _1(t)$, we have
   \bea
   2\delta _1(t)=C_4 t^{r_1}-C_1 t^n++C_5 t^{r_2}+C\frac{ t^{1-2 n}}{3 n-1}+C_4\frac{H_0
   t^{r_1+1}}{-n+r_1+1}+C_5\frac{
   H_0 t^{r_2+1}}{-n+r_2+1}.
   \eea
The arbitrary integration constants $C_4$ and $C_5$ can be determined again from the initial conditions $\delta \left(t_0\right)=\delta _0$, and $\dot{\delta }\left(t_0\right)=H_0\left(t_0\right)\delta _0$, respectively.

\section{Observational imprints on the CMB and supernovae luminosities of the non-comoving dark matter--dark energy model}\label{obs}

In the present Section, we briefly consider the  observational implications of the non-comoving dark energy and dark matter models. The anisotropic expansion of the Universe would leave its imprints on both CMB and supernovae data. The amount of a possible anisotropy existing in the Universe can
be constrained by using the CMB and SNIa observations. Usually the present CMB data constraints can be much tighter than the constraints from the anisotropies in the luminosity--distance--redshift relationship of SNIa \cite{Tomi1}.  However, there are anisotropic models which can avoid this bound completely \cite{Tomi1}. Within these bounds, the existence of anisotropy in the Universe can provide a potential explanation of various anomalies in the cosmological observations, especially in the CMB at the largest angles.

\subsection{Effects on CMB on the non-comoving motion of dark energy and dark matter}

A possible anisotropic expansion of the Universe will leave its imprint in the CMB. To calculate the CMB spectrum
in a background geometry with a Bianchi type I metric given by Eq.~(\ref{7}) we follow the approach introduced in \cite{Tomi1}. Photons emitted by astrophysical sources at cosmological distances move in the Bianchi type I Universe along the geodesic lines, given by
\be
\frac{du^{\mu }}{d\lambda }+\Gamma ^{\mu }_{\alpha \beta}u^{\alpha }u^{\beta }=0,
\ee
where $\lambda $ is an affine parameter, and the Christoffel symbols $\Gamma ^{\mu }_{\alpha \beta}$ associated to the metric Eq.~(\ref{7}) are given by $\Gamma ^{0 }_{i i}=a_i^2H_i$, and $\Gamma ^{ i}_{0 i}=H_i$, $i=1,2,3$ (no summation upon $i$ in the Christoffel symbols).

The four-velocity $u^{\mu }=dx^{\mu }/d\lambda $ of the photons satisfy the normalization condition $u^{\mu }u_{\mu }=0$, giving $\left(u_0\right)^2=a_i^2\left(u^i\right)^2$. Consider two photons, emitted at two time interval $\left(t_0 = t_e, t_1 = t_e + \tau\right)$, where $\tau \ll t_e$. By taking the difference of the two normalization conditions for photons in the first order of approximation in $\tau $ we obtain \cite{Tomi1}
\be\label{86}
u^0\frac{d\tau }{d\lambda }=\sum_{i=1}^3{a_i\dot{a}_i\left(u^i\right)^2\tau (\lambda )}+O\left(\tau ^2\right).
\ee
By introducing the redshift $z$, defined as $1+z\left(\lambda _e\right)=\tau \left(\lambda _r\right)/\tau \left(\lambda _e\right)$, where $\tau  \left(\lambda _r\right)$ is the time difference of the received signals, with the use of Eq.~(\ref{86}) we immediately obtain
\be
\frac{d}{d\lambda }\ln(1+z)=\frac{1}{u^0}\sum_{i=1}^3{a_i\dot{a}_i\left(u^i\right)^2}.
\ee

The photon velocity components $u^i$, $i=1,2,3$, can be obtained from the photon geodesic equation of motion,
\be\label{sg}
\frac{du^i}{d\lambda }+2\frac{\dot{a}_i}{a_i}u^iu^0=0, \qquad  i=1,2,3,
\ee
giving
\be
u^i(t)=\frac{u_{0i}}{a_i^2(t)}, \qquad  i=1,2,3,
\ee
where $u_{0i}$, $i=1,2,3$ are arbitrary constants of integration. In order to determine the integration constants, we adopt the convention according to which the present day $t=t_0$ value of the scale factors is $a_i\left(t_0\right) = 1$. We denote the present day values of the photon four-velocity as $u^i\left(t_0\right) = \hat{u}^i$, $i=1,2,3$. Since one
can always reparameterize the affine parameter $\lambda $ without modifying the physical results, we can normalize the present day photon four-velocities so that $\sum_{i=1}^3\left(\hat{u}^i\right)^2= 1$. Hence one may interpret the unit vector $\hat{u}$ in terms of the angles
$\left(\hat{u}_x, \hat{u}_y, \hat{u}_z\right) = \left(\sin \theta \cos \phi, \sin \theta \sin \phi, \cos \theta\right)$, representing the present day arrival angles of the photons to the observer. Therefore, by substituting the photon velocities into the redshift equation, we obtain \cite{Tomi1}
\be
1+z\left(\hat{\vec{u}}\right)=\sqrt{\sum_{i=1}^3{\frac{\hat{u}_i^2}{a_i^2}}},
\ee
which can be rewritten as
\be
1+z\left(\hat{\vec{u}}\right)=\frac{1}{a_1}\sqrt{1+\hat{u}_y^2e_y^2+\hat{u}_z^2e_z^2},
\ee
where the eccentricities $e^2_y$ and $e_z^2$ are defined as
\be
e_y^2=\left(\frac{a_1}{a_2}\right)^2-1, \qquad  e_z^2=\left(\frac{a_1}{a_3}\right)^2-1.
\ee
Since in the non-comoving dark energy-dark matter model we have considered we have assumed that $a_1=a_2$, we obtain $e_y^2=0$, and
\be
e_z^2=\frac{2}{a}\left[\delta _1(t)-\delta _3(t)\right],
\ee
respectively.

The temperature distribution  is determined by the relation
\be
T\left(\hat{\vec{u}}\right) =\frac{T_{*}}{1+z\left(\hat{\vec{u}}\right)},
\ee
where the last scattering temperature $T_{*}$ does not depend on the direction. However, due the anisotropic nature of the geometry of the Universe,  photons
coming from different directions will be redshifted by different amounts. The spatial average $\bar{T}$ of the temperature field is obtained as $4\pi \bar{T}=\int{T\left(\hat{\vec{u}}\right)d\Omega _{\hat{\vec{u}}}}$, while the anisotropies in the temperature field are given by
\bea
\delta T\left(\hat{\vec{u}}\right)&=&1-\frac{T\left(\hat{\vec{u}}\right)}{\bar{T}}
    =1-\frac{T_{*}}{\bar{T}}\frac{1}{\sqrt{1+2\left[\delta _1\left(t_0\right)-\delta _3\left(t_0\right)\right]\cos ^2\theta}}.
\eea

The multipole spectrum $Q_l$ can be described by the coefficients in the spherical expansion of the temperature anisotropy field. The quadrupole $Q_2$ can be obtained as \cite{Tomi1}
\bea
Q_2=\frac{2}{5\sqrt{3}}\sqrt{e_z^4+e_y^4-e_z^2e_y^2}=\frac{2}{5\sqrt{3}}e_z^2
  =\frac{4}{5\sqrt{3}}\frac{1}{a}\left[\delta _1(t)-\delta _3(t)\right].
\eea

The nominal value of the best fit quadrupole measured by Planck is $[2 (2 + 1)/2\pi]C_2= 299.495 \times 10^{-12}$ \cite{2}. Therefore the recent observational results can be used to constrain the parameters of the non-comoving dark energy--dark matter model.

\subsection{Constraints on non-comoving dark matter and dark energy from supernova observations}

Anisotropy in the expansion rates of the Universe would, in principle,  determine  rotationally non-invariant
luminosity distance--redshift relationships for the Ia type supernovae. The luminosity--redshift relationship of the SNIa can be used to probe the possible anisotropies in the expansion history of the Universe \cite{Tomi1, anl}. The spatial geodesic Eq.~(\ref{sg}) shows that the direction  of a photon $\hat{\vec{u}}$ that is coming towards us from an astrophysical source is constant. This follows from the orthogonality property of the Bianchi I models. The null geodesic equation can be rewritten as
\be
−dt^2 =\left(\hat{u}_x^2a_1^2+\hat{u}_y^2a_2^2+\hat{u}_z^2a_3^2\right)dr^2,
\ee
where $r^2=x^2+y^2+z^2$. The conformal distance can be obtained as $\int{dr}$.
 The luminosity distance at the redshift $z$ in the direction $\hat{u}$ is thus given by \cite{Tomi1,anl}
 \be
 d_L\left(z, \hat{u}\right) = (1 + z)\int_{t_0}^{t(z)}{\frac{dt}{\sqrt{\hat{u}_x^2a_1^2+\hat{u}_y^2a_2^2+\hat{u}_z^2a_3^2}}}.
 \ee
Therefore high precision  SNIa data can be used to constrain the non-comoving dark energy--dark matter models.

\section{Discussions and final remarks}\label{sect4}

In the present paper, we have considered the theoretical possibility that our Universe may be modeled as a mixture of two non-interacting perfect fluids, namely, dark energy and dark matter, respectively, with different four-velocities. This model is formally equivalent to a Universe filled with a  single anisotropic fluid, with different pressure components along the coordinate axes. The gravitational equations describing the non-comoving dark energy and dark matter mixture were explicitly obtained for a Bianchi type I geometry. By assuming that the difference in the four-velocities of the dark energy and dark matter is small, one can consider that the deviations from isotropy are also small, representing just a small perturbation  of the homogeneous and isotropic background metric. In particular, the perturbation equations have been explicitly obtained and solved, for the case of the Bianchi type I geometry. Two simple cases, corresponding to an exponential and power law expansion of the background geometry have also been analyzed in detail.

Therefore, if there is a slight difference between the four-velocities of the dark energy and dark matter, the Universe would acquire some anisotropic characteristics, and its geometry will deviate from the standard FLRW one. In fact, the recent Planck results show that the presence of an intrinsic large scale anisotropy in the Universe cannot be excluded {\it a priori}, so that the model presented in this work can be considered as a plausible and viable working hypothesis. However, in the present paper we have considered that the deviations from isotropy are small, and we have restricted our analysis to the linear perturbation regime. If the amplitude of the perturbations does not satisfy the condition $\delta _i \ll a$, $i=1,2,3$, then non-linear effects should also be taken into account, and incorporated into the description of the anisotropic effects generated by the non-comoving dark energy and dark matter. In the standard cosmological approach, in which dark energy and dark matter move with the same four - velocity, in the comoving frame there is no redshift effect on the common velocity. But in the present approach, with two distinct four-velocities for each component, and with the cosmological dynamics explicitly depending  on them, the dark energy fluid velocity could redshift away during the dark energy domination phase.

We have also pointed out some possibilities of observationally testing the model, by using the recently released Planck data \cite{1}, as well as the type Ia supernovae data \cite{acc}. In the case of the CMB we have explicitly obtained the distribution of the temperature anisotropy and of the quadrupole $C_2$ as functions of the deviations $\delta _1$ and $\delta _3$ from the isotropic flat geometry. Once the form of the background FLRW geometry, more exactly the form of the scale factor $a(t)$ is known, one can use the obtained relations to put tight constraints on the parameters of the non-comoving dark energy--dark matter model. Similar constraints can be obtained from the study of the anisotropic luminosity distance by using the type Ia supernovae observational data.

In this context, we emphasize that a major challenge for cosmology is to elucidate the nature of dark energy, which drives the accelerated expansion of the Universe. The simplest explanation for dark energy is the cosmological constant, where the latter has an equation of state $w=p_{DE}/\rho _{DE}=-1$. Scalar fields, alternatives to the cosmological constant, have an equation of state with $w\geq -1$. However, the CMB data alone does not strongly constrain $w$, due to the two-dimensional geometric degeneracy in the theoretical models. The recent results of Planck, combined with the supernova data, provide $w=-1.09\pm 0.17$, or $w=-1.13^{+0.13}_{-0.14}$, generally in a good agreement with the cosmological constant hypothesis. However, at the present level of observational precision one cannot rule out the existence of a dark energy component of the Universe, distinct from a cosmological constant. For the case of a ``pure'' cosmological constant, the formalism developed in the present paper cannot be applied, since there is no flow associated to $\Lambda $.

In the case of the scalar field $\phi $, with energy-momentum tensor
\be
T_{\mu \nu}^{(\phi)}=\nabla _{\mu }\phi \nabla _{\nu }\phi -g_{\mu \nu}\nabla ^{\lambda }\phi \nabla _{\lambda }\phi /2-V(\phi )g_{\mu \nu},
 \ee
 where $V(\phi)$ is the scalar field potential, a perfect fluid description corresponding to an energy-momentum tensor $T_{DE\;\mu \nu}=\left(\rho _{\phi}+p_{\phi}\right)u_{\phi \;\mu}u_{\phi \:\nu }-p_{\phi }g_{\mu \nu}$ can be obtained, if $\nabla ^{\lambda }\phi $ is timelike, by introducing the effective four-velocity of the scalar field fluid as \cite{Fa}
\be
u_{\phi \mu}=\frac{\nabla _{\mu}}{\sqrt{\left|\nabla ^{\lambda }\phi \nabla _{\lambda }\phi \right|}},
\ee
where $\nabla ^{\lambda }\phi \nabla _{\lambda }\phi\neq 0$, and $u_{\phi \mu}u^{\mu }_{\phi}={\rm sign}\left(\nabla ^{\lambda }\phi \nabla _{\lambda }\phi\right)$. The effective fluid pressure and energy density associated to the scalar field energy momentum tensor can be obtained as $\rho _{\phi}=T_{\mu \nu}^{(\phi)}u^{\mu }_{\phi}u^{\nu}_{\phi}$ and $p_{\phi }=T_{\mu \nu}^{(\phi)}h^{\mu \nu}/3$, where $h^{\mu \nu}=g^{\mu \nu}-u^{\mu }_{\phi}u^{\nu}_{\phi}$, and are given by \cite{Fa}
\be
\rho _{\phi}=\left[\nabla ^{\lambda }\phi \nabla _{\lambda }\phi/2-V(\phi)\right]{\rm sign}\left(\nabla ^{\lambda }\phi \nabla _{\lambda }\phi\right),
 \ee
 and
 \bea
 p_{\phi }&=&\left[-1+{\rm sign}\left(\nabla ^{\lambda }\phi \nabla _{\lambda }\phi\right)/2\right] \nabla ^{\lambda }\phi \nabla _{\lambda }\phi\nonumber\\
 &&-\left[4+{\rm sign}\left(\nabla ^{\lambda }\phi \nabla _{\lambda }\phi\right)\right]V(\phi),
 \eea
 respectively. Therefore if the four-velocities of the dark matter fields, $u_{DM}^{\mu }$ and of the scalar field $u_{\phi}^{\mu }$ satisfy the condition
 \be
 u_{DM}^{\mu }\neq u_{\phi}^{\mu },
 \ee
 the mixture of the non-comoving scalar field (dark energy) and dark matter can be described by a single anisotropic fluid. Such a situation may have occurred during the inflationary, or post-inflationary (reheating)  era of the evolution  of the Universe.

 In the present paper we have investigated only homogeneous cosmological models, favored by present day observations,  in which $u_{DM}$, $u_{DE}$, $\rho _{DM}$, $p_{DM}$ and $p_{DE}$  depend on time only. The possible dependence on the spatial coordinates $\left(x^1, x^2, x^3\right)$ of the physical and geometrical quantities, and the symmetries induced by this dependence, have  not been considered. There is strong observational evidence (see the recent Planck data \cite{1}-\cite{5}) that on the large scale the Universe is homogeneous, with possible small deviations from isotropy. The present model can be easily generalized for the case of a three component mixture, or for cases involving the presence of condensates and multi-component scalar fields. These possibilities, their impact on the cosmological evolution, as well as some observational implications, will be considered in  future studies.

\section*{Acknowledgments}

We would like to thank to the anonymous referee, whose comments and suggestions helped us to significantly improve our manuscript. FSNL acknowledges financial support of
the Funda\c{c}\~{a}o para a Ci\^{e}ncia e Tecnologia through the
grants CERN/FP/123615/2011 and CERN/FP/123618/2011.

\end{document}